\documentclass[12pt, a4paper]{article}  
\usepackage{arxiv}
\usepackage{amsmath,amsfonts,amssymb}
\usepackage{graphicx}
\usepackage[colorlinks=true, allcolors=blue]{hyperref}
\usepackage{threeparttable} 
\usepackage{macro_ref}
\usepackage{geometry}
 \usepackage{natbib}

\title{GRIP: a generic data reduction package for nulling interferometry}

\renewcommand{\undertitle}{}
\renewcommand{\headeright}{}
\renewcommand{\shorttitle}{GRIP}
\date{}

\begin{document} 
\maketitle

\begin{center}
    Marc-Antoine Martinod$^a$\footnote{mam.astro01@gmail.com}, 
    Denis Defrère$^a$,
    Romain Laugier$^a$,
    Steve Ertel$^{b,c}$,
    Olivier Absil$^d$,
    Barnaby Norris$^e$,
    Bertrand Mennesson$^f$.
    ~\newline
    ~\newline
    $^a$Institute of Astronomy, KU Leuven, Celestijnenlaan 200D, 3001 Leuven, Belgium;
    $^b$Department of Astronomy and Steward Observatory, 933 North Cherry Ave, Tucson, AZ 89 85721, USA;
    $^c$Large Binocular Telescope Observatory, 933 North Cherry Ave, Tucson, AZ 85721, USA;
    $^d$Space sciences, Technologies \& Astrophysics Research (STAR) Institute, University of Li\`ege, Li\`ege, Belgium;
    $^e$Sydney Institute for Astronomy, School of Physics, Physics Road, University of Sydney, NSW 2006, Australia;
    $^f$Jet Propulsion Laboratory, California Institute of Technology (United States).
\end{center}




\begin{abstract}
Nulling interferometry is a powerful observing technique to study exoplanets and circumstellar dust at separations too small for direct imaging with single-dish telescopes. 
With recent photonics developments and the near-future ground-based instrumental projects, it bears the potential to detect young giant planets near the snow lines of their host stars.  
The observable quantity of a nulling interferometer is called the null depth, its precise measurement and calibration remain challenging against instrument and atmospheric noise.
Null self-calibration is a method aiming to model the statistical distribution of the nulled signal. 
It has proven to be more sensitive and accurate than average-based data reduction methods in nulling interferometry.
The variety of existing and upcoming of nullers raises the issue of consistency of the calibration process, structure of the data and the ability to reduce archived data on the long term.
It has also led to many different implementations of the Null self-calibration method.
In this article, we introduce \texttt{GRIP}: the first open-source toolbox to reduce nulling data with enhanced statistical self-calibration methods from any nulling interferometric instrument within a single and consistent framework.
Astrophysical results show good consistency with two published GLINT and LBTI datasets and confirm nulling precision down to a few 10$^{-4}$.

\end{abstract}


\section{Introduction}
Nulling interferometry is a powerful technique to perform high-contrast imaging in the close neighbourhood of the stars \cite{bracewell1978}.
Pioneered on the Multiple-Mirror Telescope (MMT) in 1998 \cite{Hinz1998}, it has been used over the past decades to detect and characterize habitable-zone dust around nearby stars.
Similarly to the coronagraphy technique in direct imaging, nulling interferometry suppresses the on-axis source (e.g. a star) but proceeds differently : beams from two apertures are coherently combined to create a destructive interference for on-axis sources.
The offset in the optical path lengths induced by an off-axis sources (e.g.\ an exoplanet) is transmitted by the instrument, allowing a direct detection and characterisation of the object.
The nulled depth $N$ is linked to the modulus of the observed target's spatial coherence (also known as the visibility) $V$ with $N = \frac{1-V}{1+V}$.
Performing nulling requires extremely stable interferences to efficiently discriminate the photons coming from the planets from the ones coming from the starlight.
The fainter the off-axis source, the better the stabilisation must be, to efficiently remove the starlight from the measured signal.
Such observation is challenged by atmospheric turbulence and vibrations in the instrument and telescope.

Legacy data reduction techniques relied on measuring the average value of the null depth on the science and calibrator targets then performed a subtraction between the two to obtain the astrophysical null depth, solely governed by the object's geometry.
The state-of-the-art method, to process nulling data sets, is the nulling self-calibration (NSC) \cite{hanot2011}: it models the statistical fluctuations to provide a self-calibrated astrophysical null depth.
In other words, this method directly measures the astrophysical null depth by fitting out null biases related to the instrumental response of the nulling interferometer.
Therefore, the measurement is more precise and the observation of a reference star to measure this response is ideally not required.

The null self-calibration is a simulation-based inference method \cite{Cranmer2020}.
Monte-Carlo simulations are carried out to reproduce the measured distributions.
The model takes for parameters the distributions of the instrumental errors and the astrophysical null depth.
The outcome is a sequence of simulated nulled flux.
These data are sorted into a histogram, which is then compared with the histogram of the observed data through the optimization of a simple cost function.
This process enables the self-calibration, provided that the simulator can correctly model the dominant effects of the instrument response. 
In some cases, calibration with reference targets is still required to remove sources of null errors that have not been simulated.
One of the parameters is the astrophysical null depth, i.e. the observable which only depends on the surface brightness and the geometry of the target, free from the bias of the instrument response.
The use of this technique on the Palomar Fiber Nuller yielded an improvement on the precision by a factor of 10 \cite{hanot2011, Mennesson2011, serabyn2019}.
This method is now used on active nullers such as the Large Binocular Telescope Interferometer (LBTI) \cite{defrere2016, Mennesson2016} and  Guided-Light Interferometric Nulling Technology (GLINT) \cite{norris2020, martinod2021}, yet through different implementations given the constraints from the instrument designs and how well characterised they were (e.g. grid search for the LBTI, least squares estimation for GLINT).
It has enabled several scientific achievements on spectroscopic binaries \cite{serabyn2019}, exozodiacal dust \cite{defrere2015, defrere2021, Ertel2018, Ertel2020} and stellar diameters \cite{martinod2021}.

By relying on a Monte-Carlo approach and a clearly defined instrumental function, this technique can be used on any nuller.
It shifts, for the first time, the way of doing data reduction in interferometry from a tailor-made pipeline for each instrument to a generic one providing self-calibrated measurements.
This paper introduces the \emph{Generic data reduction for nulling interferometry package} (GRIP\cite{martinod2024_spie}), the first project which aims to provide a collection of functions to reduce data from all existing (e.g. GLINT \cite{martinod2021} or LBTI \cite{defrere2016}) and future nulling instruments (e.g. Asgard/NOTT \cite{defrere_path_2018} and the LIFE space mission \cite{Quanz2018, Quanz2022}) in a consistent framework. 
This work is performed within the framework of the Asgard/NOTT project \cite{defrere_path_2018,defrere_hi-5_2018,defrere_l-band_2022}, a future nulling interferometer under construction for the visitor focus of the VLTI .

\section{Concept and workflow of GRIP}
Having several nullers in different places of the world that observe at different wavelengths and rely on different architectures is an advantage to grasp a detailed knowledge of a given target.
However, this represents a challenge for the consistency of the calibration process, different for each instrument, the structure of the data and the capability to reduce archived data in the future.
A data reduction pipeline built upon the NSC technique tackles these challenges as it provides a unique, consistent, easy to maintain and universal way to reduce data from any kind of nuller.
For instance, the \texttt{Vortex Imaging Processing} package \cite{gonzalez2017, Christiaens2023JOSS} provides a collection of image processing routines that can be used to create pipelines to reduce direct imaging data \cite{Christiaens2023, Christiaens2024}.

\texttt{GRIP} is the first project in nulling interferometry providing a unique and consistent framework to provide self-calibrated null depth measurements.
This toolbox offers several battle-tested optimisation strategies and cost functions to achieve this task.
\texttt{GRIP} is a package designed to build a data reduction pipeline; it does not come as a stand-alone software.

\begin{figure}[hbtp]
    \centering
    \includegraphics[width=0.8\textwidth]{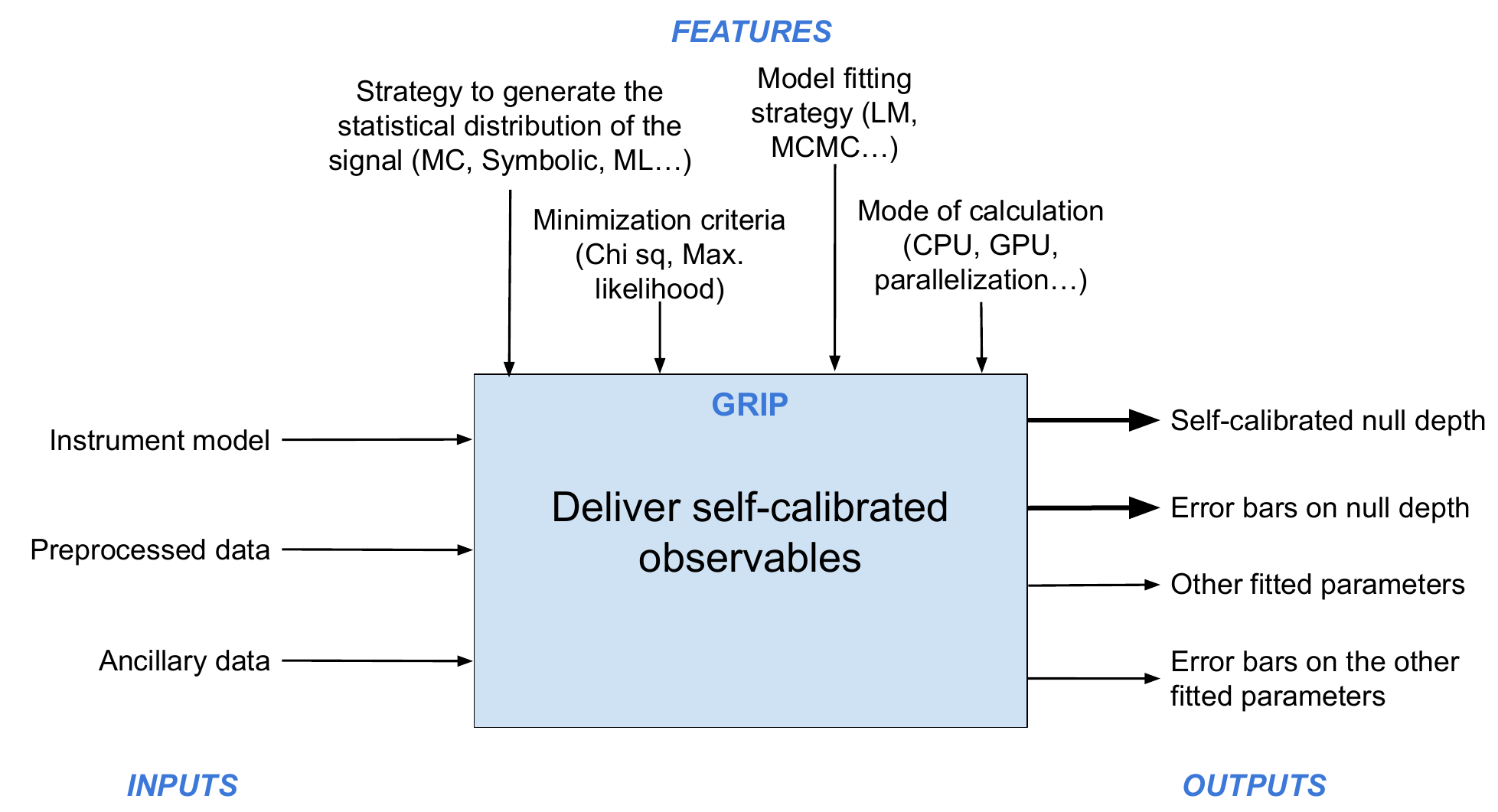}
    \caption{Structured analysis and design technique of \texttt{GRIP}. The left part lists the elements provided by the user and the right parts are the outputs of \texttt{GRIP}. The upper side lists the features of \texttt{GRIP} to process the data.}
    \label{fig:grip_sadt}
\end{figure}

\begin{figure}[hbtp]
    \centering
    \includegraphics[width=\textwidth]{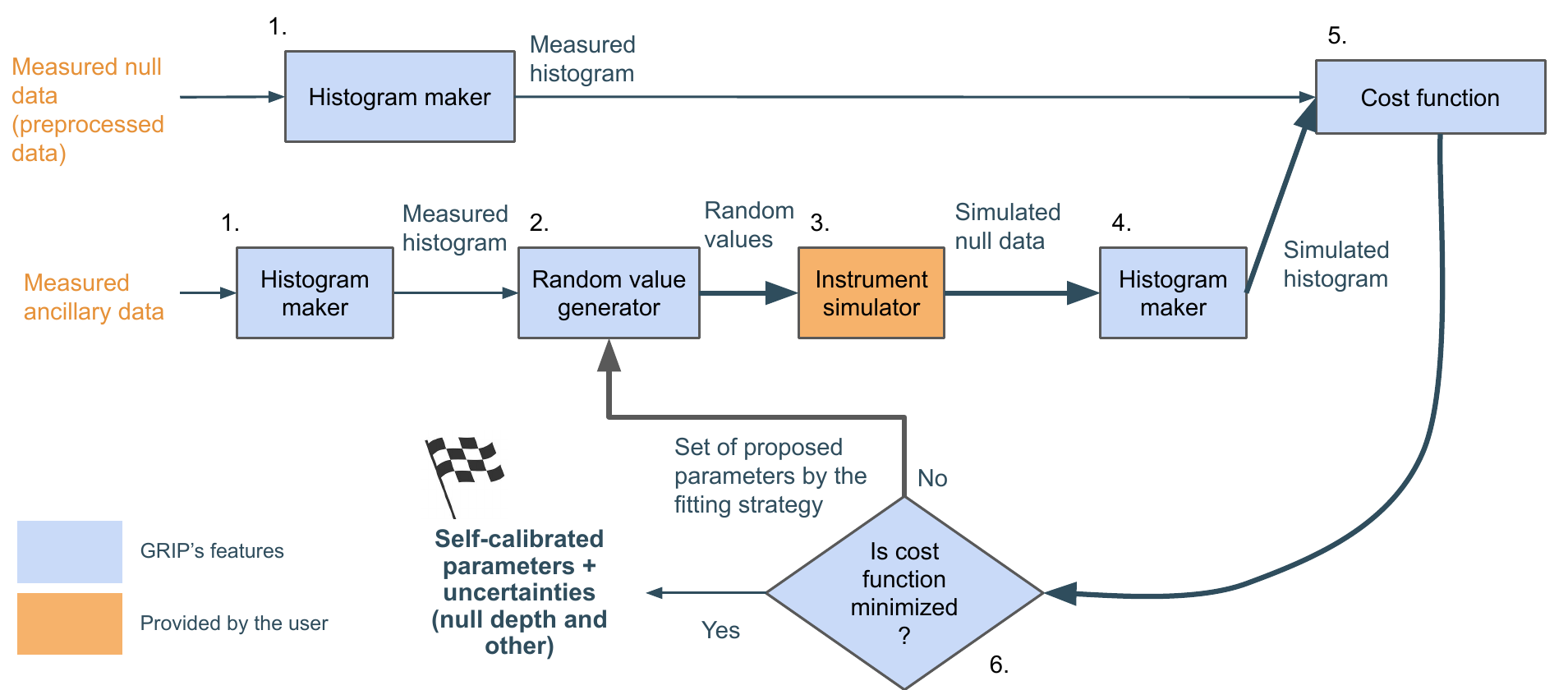}
    \caption{Workflow of GRIP. The blue boxes are GRIP's features and the blue text represents items generated by them. The orange box and text represents elements brought by the user.}
    \label{fig:workflow}
\end{figure}

Figure~\ref{fig:grip_sadt} details the inputs, outputs and features of GRIP to perform self-calibration.
In the inputs, the user has to provide a simulator of the instrument (also called \textit{simulator}), preprocessed data to extract the astrophysical null depth and ancillary data to be used by the simulator.
The preprocessed data consists of sequences such as the flux or counts of outputs.
Ancillary data are non-nulled data (e.g. photometries of the beams) and noise measurements (e.g. detector noise, thermal background, phase fluctuations...) needed to simulate the nulled signal.
\texttt{GRIP} provides tools to turn these sequences of nulled and ancillary data into histograms.
With the instrument model, it enables the generation of sequences of values of the same kind as the preprocessed data.
These simulated sequences will then be turned into histograms.
\texttt{GRIP}'s features (such as optimisation strategies and cost functions) enable the processing of the histograms from the preprocessed and simulated data to deliver the self-calibrated quantities and the other fitted parameters.
The workflow of GRIP (Fig.~\ref{fig:workflow}) follows these steps:
\begin{enumerate}
    \item The user provides the instrument response function and identifies the free parameters to be fitted.
    \item The measured null depth data (from the preprocessed data) and the measured ancillary are turned into histograms with \texttt{GRIP};
    \item The measured histograms of the ancillary data are used as probability density distributions to simulate their corresponding instrumental/atmospheric noises with the model;
    \item Random sequences generated from the empirical distributions from the previous step are injected into the instrument simulator (provided by the user) to simulate nulled signal;
    \item The simulated sequence is turned into histograms with \texttt{GRIP};
    \item The two histograms are used to calculate a cost function;
    \item The cost function is minimized in a fitting algorithm (e.g. Levenberg Marquardt, MCMC, or other) until convergence by repeating the last 3 steps. The values of the free parameters at the convergence are the self-calibrated quantities.
\end{enumerate}

\section{Package overview}
The code is being developed in \texttt{Python}, under the MIT license, and it is published on \texttt{Github}\footnote{\url{https://github.com/mamartinod/grip}}.
Its dependencies are widely-used Python packages such as \texttt{NumPy} and \texttt{scipy}.
\texttt{GRIP} is a function-based package to give flexibility to the users to implement it into their pipeline.
The documentation is made of description of the functions and tutorials, explaining the use of the features of the package.
Particularly, one tutorial is provided to help the design of an instrument simulator to use with \texttt{GRIP}.
\texttt{GRIP} runs on a GPU with the use of \texttt{Cupy} \cite{cupy_learningsys2017} or on CPU if this package is not installed when the package is used.

\texttt{GRIP}'s features can be used to build a data reduction pipeline.
These features are set in several modules (Table~\ref{tab:grip_modules}).
A Monte-Carlo process is used to generate a sequence of data, turned into a histogram, which is fit to the observed histogram using optimisers from well-known third-party Python packages.
The user defines the quantities to be generated during the Monte Carlo process and \texttt{GRIP} can efficiently reproduce any arbitrary distribution, allowing the use of the distributions of measured data.

\begin{table}[h]
    \centering
    \caption{Modules of GRIP}
    \begin{tabular}{p{4cm}p{12cm}}
    \hline
    \hline
        \texttt{fitting} & Fitting optimisers, estimators for fitting and a parameter grid search \\
        \texttt{generic} & Miscellaneous functions \\
        \texttt{histogram\_tools} & Histograms makers and random values generators \\
        \texttt{instrument\_models} & Instrument models such as GLINT \cite{martinod2021} or LBTI \cite{defrere2016} \\
        \texttt{load\_files} & Load HDF5 and FITS files \\
        \texttt{plots} & Plot results of GRIP \\
        \texttt{preprocessing} & Features such as sorting and data binning \\
    \hline
    \end{tabular}
    \label{tab:grip_modules}
\end{table}

\texttt{GRIP} is made of several modules, providing all the tools to make histograms, perform self-calibration measurements, visualise the results and check the goodness of the fit.
The \texttt{fitting} and \texttt{histogram\_tools} are core modules of \texttt{GRIP}.
They contain all the utilities to perform null self-calibration (optimiser, cost function, random value generator, histogram maker) and can handle user-defined instrument simulators.
The current optimisers available in GRIP are:
\begin{itemize}
    \item \texttt{scipy.optimize.least\_squares} with the dogbox method (least squares method);
    \item \texttt{scipy.optimize.minimize} with the Powell method (maximum likelihood estimator, or MLE);
    \item MCMC powered with the \texttt{emcee} package \cite{ForemanMackey2013} (Bayesian inference method). The used sampler is a mixture of  \texttt{StretchMove} and  \texttt{WalkMove} with the respective relative weight 0.2 and 0.8. This mixture minimizes the autocorrelation time compared to the use of a single sampler.
\end{itemize}
All methods accept bounded parameter space.
The user can define these bounds with the help of the search grid to spot the optimal region to investigate with a fitting strategy.
The sampled posterior by the MCMC method is the product of a prior function and the likelihood function.
While the user can choose the likelihood function, the prior function is a uniform distribution with customisable boundaries.
Alongside optimisers and likelihood estimators, \texttt{fitting} includes a parameter space scanner (\texttt{fitting.explore\_parameter\_space}).
This feature allows the exploration of the parameter space in a brute-force manner for the user to define the appropriate boundaries for the model fitting.

Several likelihood estimators are provided:
\begin{itemize}
    \item the logarithm of the binomial distribution: $ f(x, \theta) = \frac{n!}{x_1! \cdots x_k!} p_1(\theta)^{x_1} \cdots p_k(\theta)^{x_k} $, with $n$ the total number of elements in the sequence, $x_k$ and $p_k$ are respectively the number of elements and the probability of an element to fall into in the $k$-th bin, and $\theta$ is the vector of parameters;
    \item the sum of the weighted squared residuals: $ f(\theta) = \sum_k \frac{(d_k - p_k)^2}{\sigma_k^2}$, with $d_k$ the measured histogram, $p_k$ the model of the histogram, $\sigma_k$ is the standard deviation of $d_k$ and $\theta$ is the vector of parameters;
    \item the Pearson's $\chi^2$ \cite{Pearson1900}: $\chi^2 = \sum_k \frac{(d_k - N p_k(\theta))^2}{N p_k(\theta)}$, with $d_k$ the measured histogram, $p_k$ the model of the histogram at the k-th bin, $N$ the number of events filling the histogram and $\theta$ is the vector of parameters.
\end{itemize}
Some of the nullers are already implemented in  \texttt{instrument\_models} such as GLINT \cite{martinod2021} or LBTI \cite{Hinz2016, Ertel2020}.

Functions in \texttt{histogram\_tools} create histograms from measured data or simulations, perform random values generation and deliver diagnostic data to help to assess the reliability of the fit.
The module \texttt{plots} includes functions to display the results of the fit, the parameter space as well as the diagnostic data.
The module \texttt{load\_files} includes functions to read HDF5 and FITS files; the user specifies the keywords to read, and the functions return a dictionary with the requested data.
These can be then preprocessed with the tools, such as frame binning, sorting, slicing, and extracting spectral or photometric information, present in the \texttt{preprocessing} module.

\section{Study cases of GRIP with real data}
\texttt{GRIP} has been successfully used to measured null depth on already-published data from two instruments: GLINT and the LBTI nuller, with different features (Tab.~\ref{tab:case_study}).
This library has successfully processed data on Arcturus \cite{martinod2021} taken with GLINT and on $\beta$~Leo \cite{defrere2016} taken with the LBTI.
For both, the same parameters were retrieved: the astrophysical null depth, and the average and standard deviation of a normal distribution modelling the phase fluctuations.
\texttt{GRIP} used the same models found in their respective articles.
The cases of study are conducted on two strategies implemented in \texttt{GRIP}: the MLE and the MCMC.

\begin{table}[htbp]
    \centering
    \caption{Main features of GLINT and LBTI nullers}
    \begin{tabular}{p{5cm}|p{5cm}p{5cm}}
        \hline
        \hline
        Features & GLINT & LBTI \\
        \hline
         Spectral band & $1.5 \pm 0.5~\mu$m & $11 \pm ~2.6\mu$m\\ 
         Spectral resolution & 160 & Broadband\\
         Combiner platform & Integrated-optics & Bulk optics\\
         AO correction & Yes & Yes\\
         Fringe tracking & No & Yes\\
         Acquisition of photometry, constructive and destructive interferences & Simultaneous & Sequential\\
         Limiting noises & Phase fluctuations and camera read-out & Phase fluctuations and thermal background\\
        \hline
    \end{tabular}
    \label{tab:case_study}
\end{table}

\subsection{Self-calibration with Maximum Likelihood Estimators}
\begin{figure}[h]
    \centering
    \begin{tabular}{c}
        \includegraphics[width=0.8\textwidth]{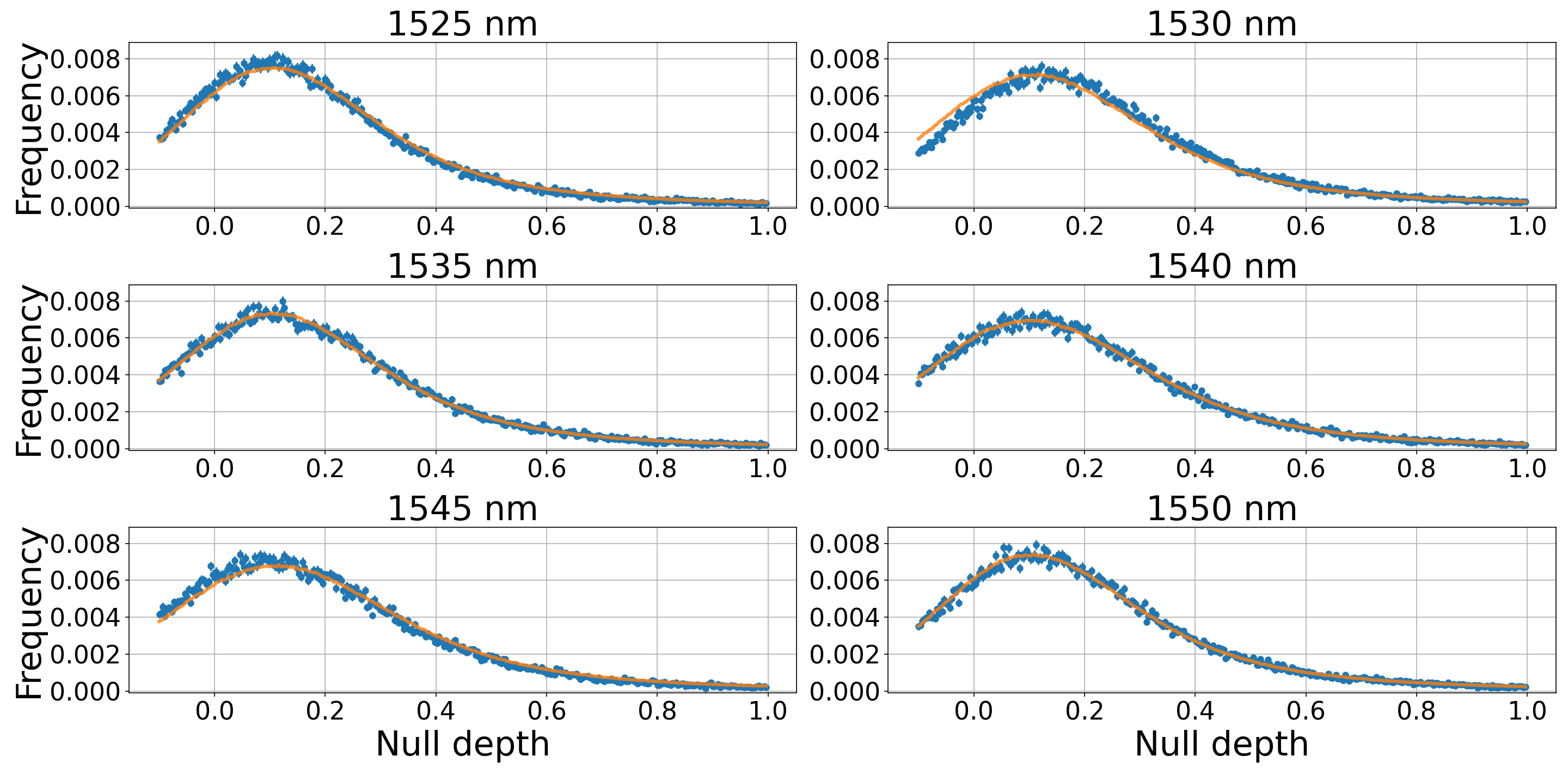} \\ 
        \includegraphics[width=0.8\textwidth]{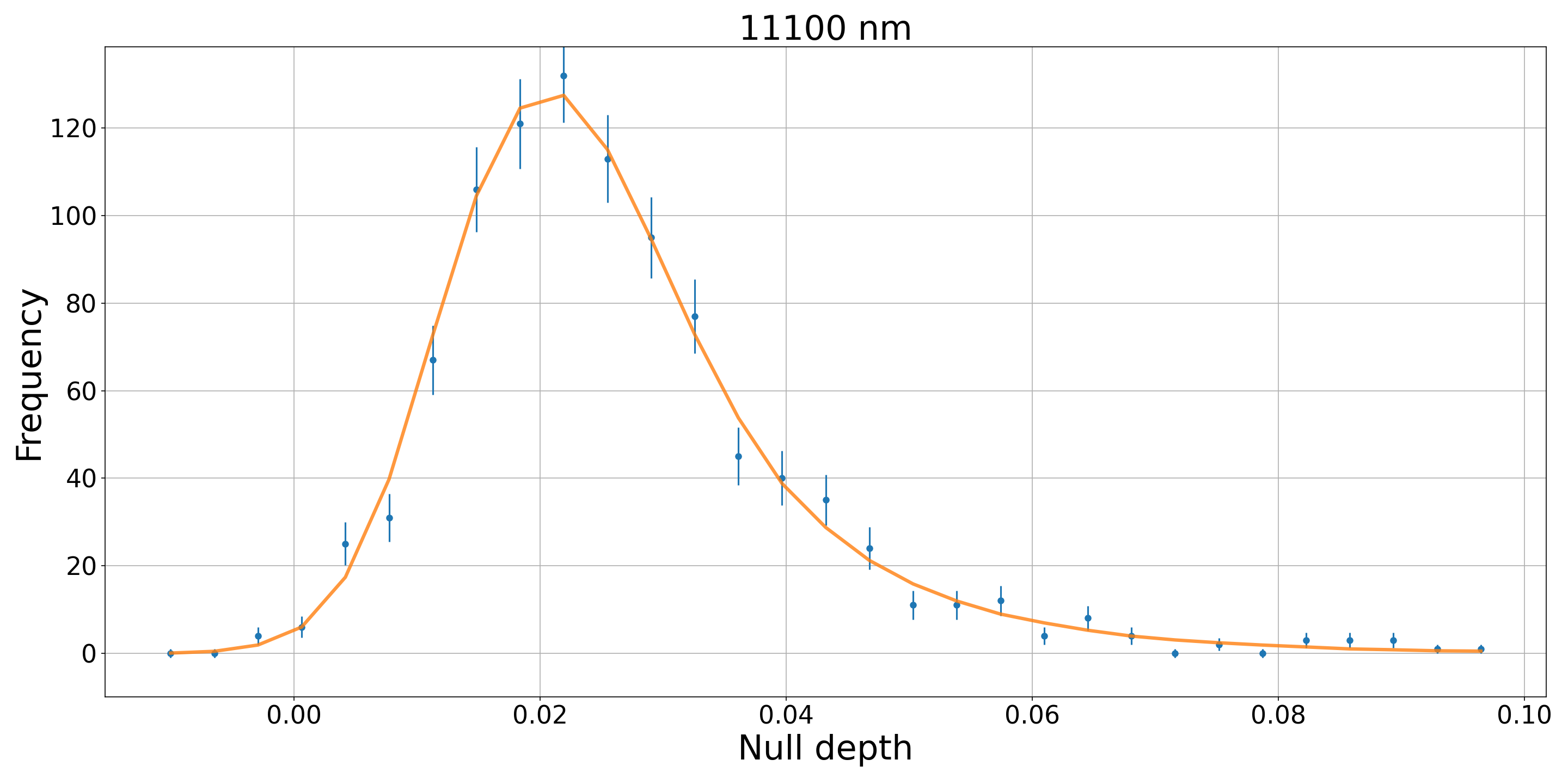} \\
    \end{tabular}
    \caption{Fitted histograms of GLINT (top) and LBTI (bottom) data with \texttt{GRIP}. 
    Top: GLINT data are Arcturus, taken with a baseline of 5.55 meters. The histograms of the null depth is made for every spectral channel are fitted at once. The fitted values are: $N_a = 0.0705 \pm 0.000137$, $\mu_{OPD} = 302 \pm 7.34$~nm and $\sigma_{OPD} = 163 \pm 0.378$~nm. 
    Bottom: LBTI data are one observing block of $\beta$~Leo, taken with a nominal baseline of 14.4 meters centre to centre, there is no spectral dispersion. The fitted values are: $N_a = 0.00714 \pm 0.00056$, $\mu_{OPD} = 174 \pm 11.4$~nm and $\sigma_{OPD} = 286 \pm 1.01$~nm. Unlike for GLINT data (left), the fitted null depth on this LBTI data set needs a calibration process with a reference star as the instrument simulator does not capture all the instrumental noise biases.}
    \label{fig:histos}
\end{figure}

For the GLINT case, \texttt{GRIP} uses the same cost function and fitting strategy used in \cite{martinod2021}, namely the sum of the squared residuals and \texttt{scipy.optimize.least\_squares}.
For the LBTI case, \texttt{GRIP} uses a different estimator of the binomial mass probability function from the LBTI reduction code and a different fitting strategy than what was used in \cite{defrere2016}.
The cost function comes from the widely-used package \texttt{scipy} whereas LBTI reduction code uses an internally-developed estimator.
The fitting strategy used in \texttt{GRIP} relies on the \texttt{minimize} method of \texttt{scipy} while the LBTI code performs a grid search.
Both fits are performed on null depths calculated from fluxes of the destructive interference recorded in the camera (Fig.~\ref{fig:histos}).

\subsection{Self-calibration with MCMC}
\begin{figure}[h]
    \centering
    \begin{tabular}{cc}
        \includegraphics[width=0.48\textwidth]{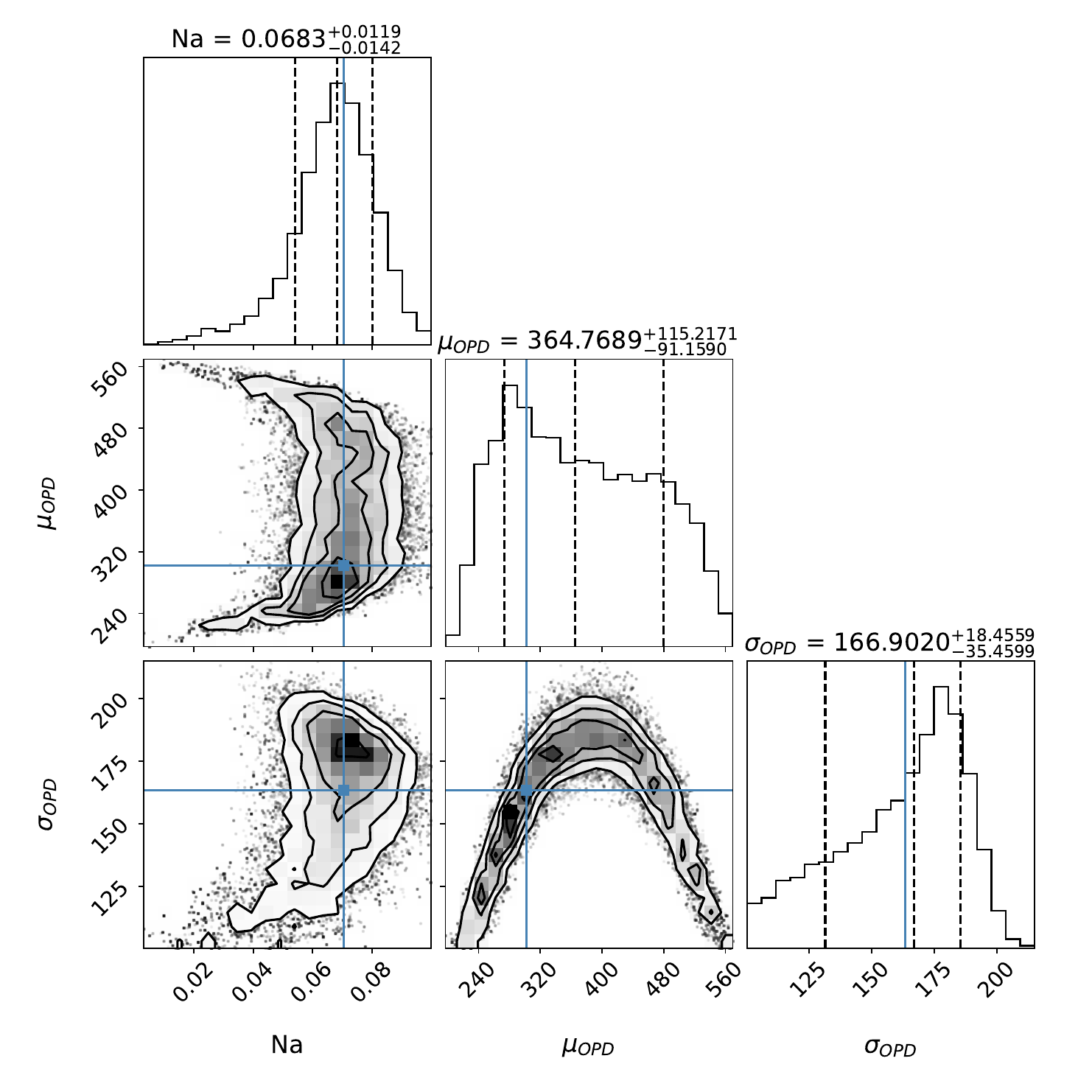} & 
        \includegraphics[width=0.48\textwidth]{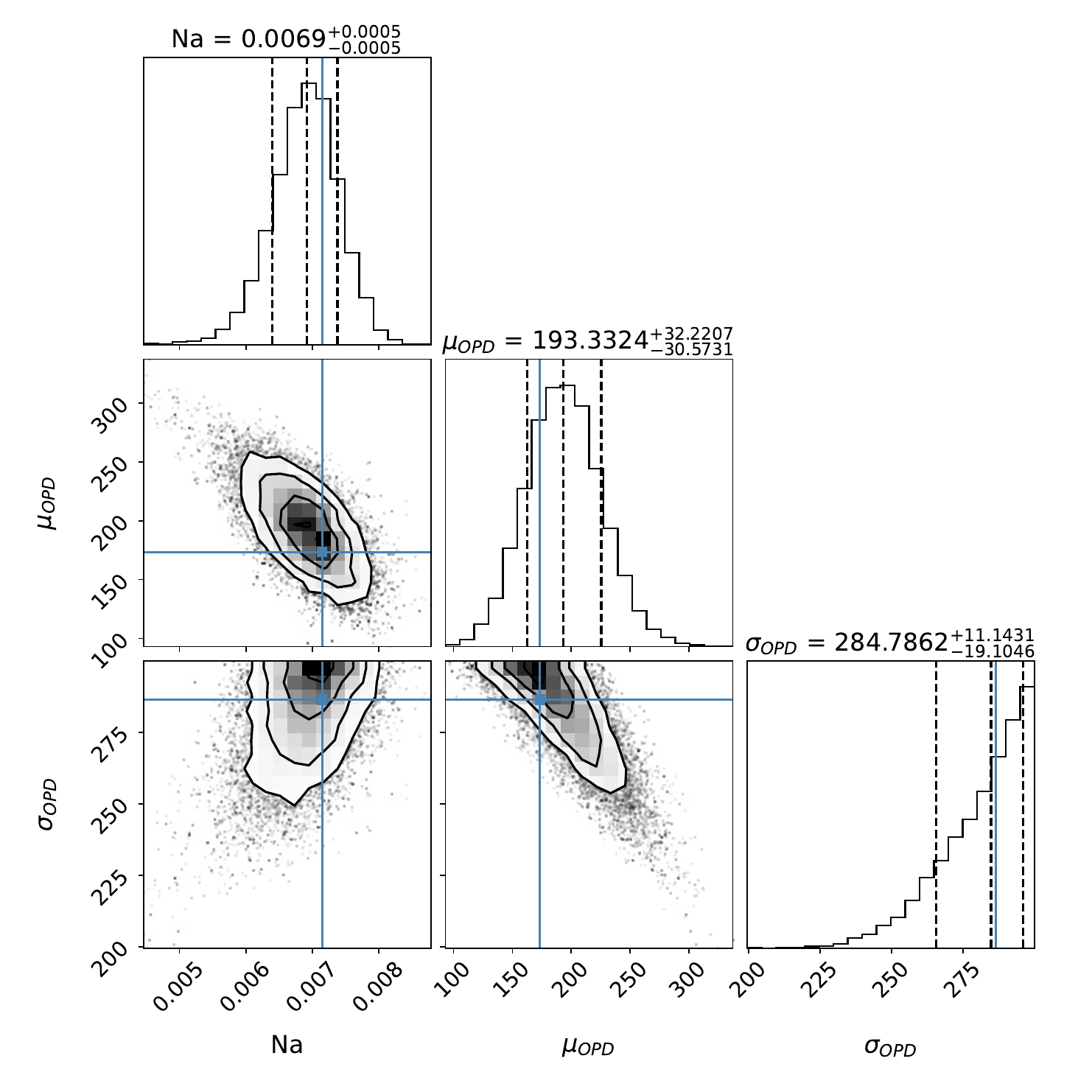} \\
    \end{tabular}
    \caption{Corner plots of the inferred parameters of GLINT (left) and LBTI (right) data. The inferred parameters are the null depth $Na$, the location $\mu_{\textrm{OPD}}$ and scale $\sigma_{\textrm{OPD}}$ parameters of the normal distribution describing the phase fluctuations, in nanometre. The blue lines represent the fitted values found with MLE method and use as the starting point for the MCMC algorithm. The values summarising the posterior distributions are the median, the 16-th and 84-th quantiles and are display on the top of each column of the corner plot. 
    Left: same GLINT data as on the figure~\ref{fig:histos}. 
    Right: same LBTI data as on the figure~\ref{fig:histos}.}
    \label{fig:mcmc}
\end{figure}

\texttt{GRIP} offers the possibility to perform MCMC with the \texttt{emcee} package.
The main advantage of using this technique over MLE is having a more representative estimation of the posterior of the parameters, at the expense of a longer process if the walkers are not initialised near the optimal solutions, compared to MLE methods.
For both GLINT and LBTI study cases, the fitted parameters are consistent with the ones found with the MLE techniques (FIgure~\ref{fig:mcmc}).

\subsection{Measuring the LBTI nuller transfer function}
\begin{figure}[htbp]
    \centering
    \includegraphics[width=0.8\textwidth]{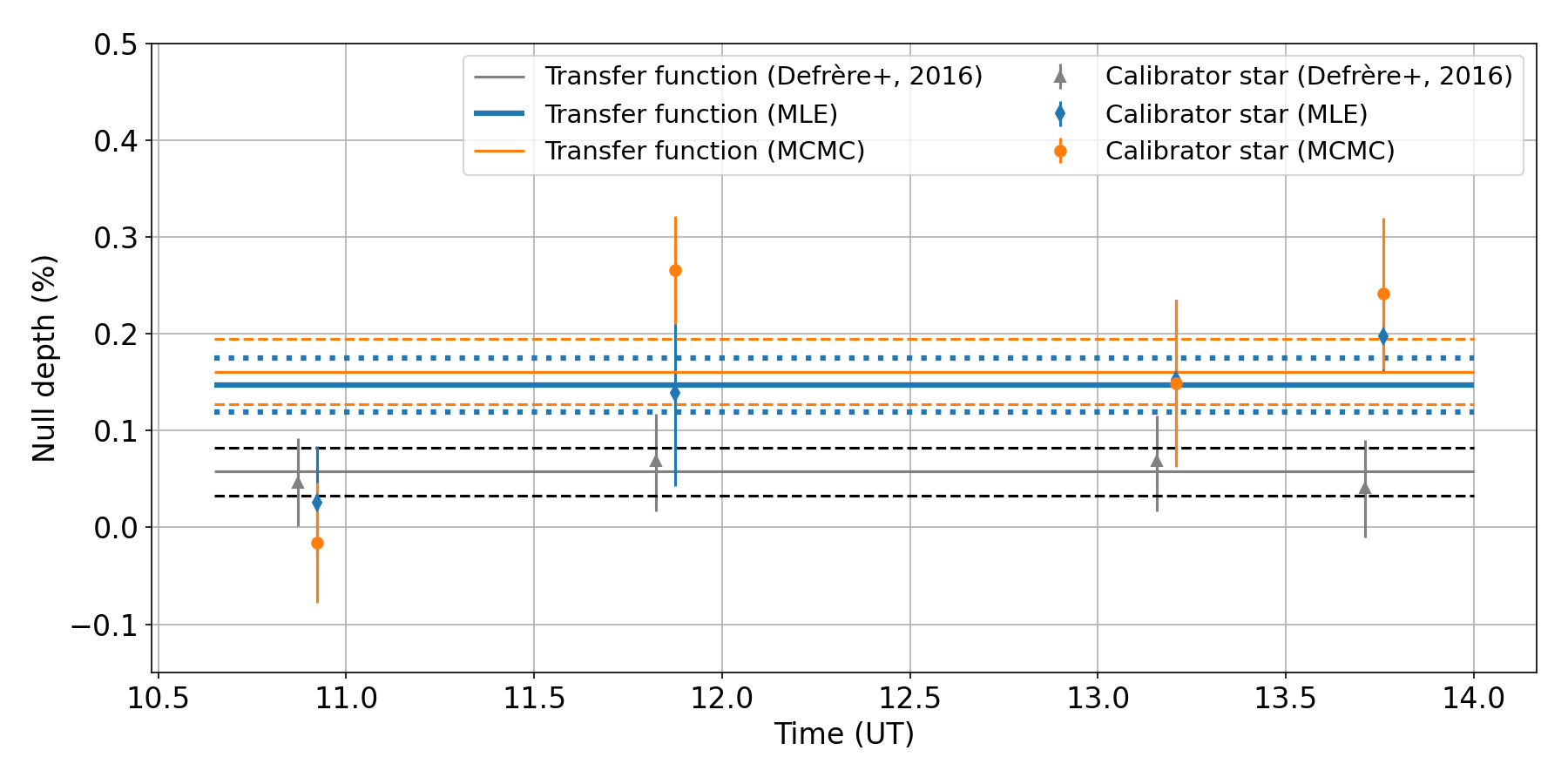}
    \caption{Transfer functions of the LBTI obtained with \texttt{GRIP} with the MLE and MCMC methods. They are compared to the published one \cite{defrere2016}.}
    \label{fig:beta_leo_ft}
\end{figure}

While the GLINT instrument model directly provides self-calibrated null depth, the model of the LBTI is not accurate enough and a calibration process is needed.
Following the procedure detailed by \cite{defrere2016}, we have reduced the full dataset, made of one science target ($\beta$~Leo) and three calibrators (HD~104979, HD~109742 and HD~108381) and calculated the transfer function of the instrument.
These steps have been performed with the outputs from \texttt{GRIP} with the MLE and MCMC methods, and compared with the one published by \cite{defrere2016} (Fig.~\ref{fig:beta_leo_ft}).
The transfer functions obtained from the MLE results is $0.147 \pm 0.028$~\% and $0.161 \pm 0.034$~\% from the MCMC results.
Both transfer function are consistent with each other while being higher than the published one.
A bias remain between \texttt{GRIP}'s transfer functions and the published one ($0.057 \pm 0.025$\%), that appears to be systematics, with little incidence on the calibration of the null depths of the science star (see Section~\ref{sec:comparison}). 
These results confirm the LBTI nulling accuracy on the transfer function of a few 10$^{-4}$, required for the exozodiacal disk survey of the LBTI \cite{Ertel2018,Ertel2020}.

\subsection{Comparison of the methods with the published results}
\label{sec:comparison}
\begin{figure}[h]
    \centering
    \begin{tabular}{c}
        \includegraphics[width=0.75\textwidth]{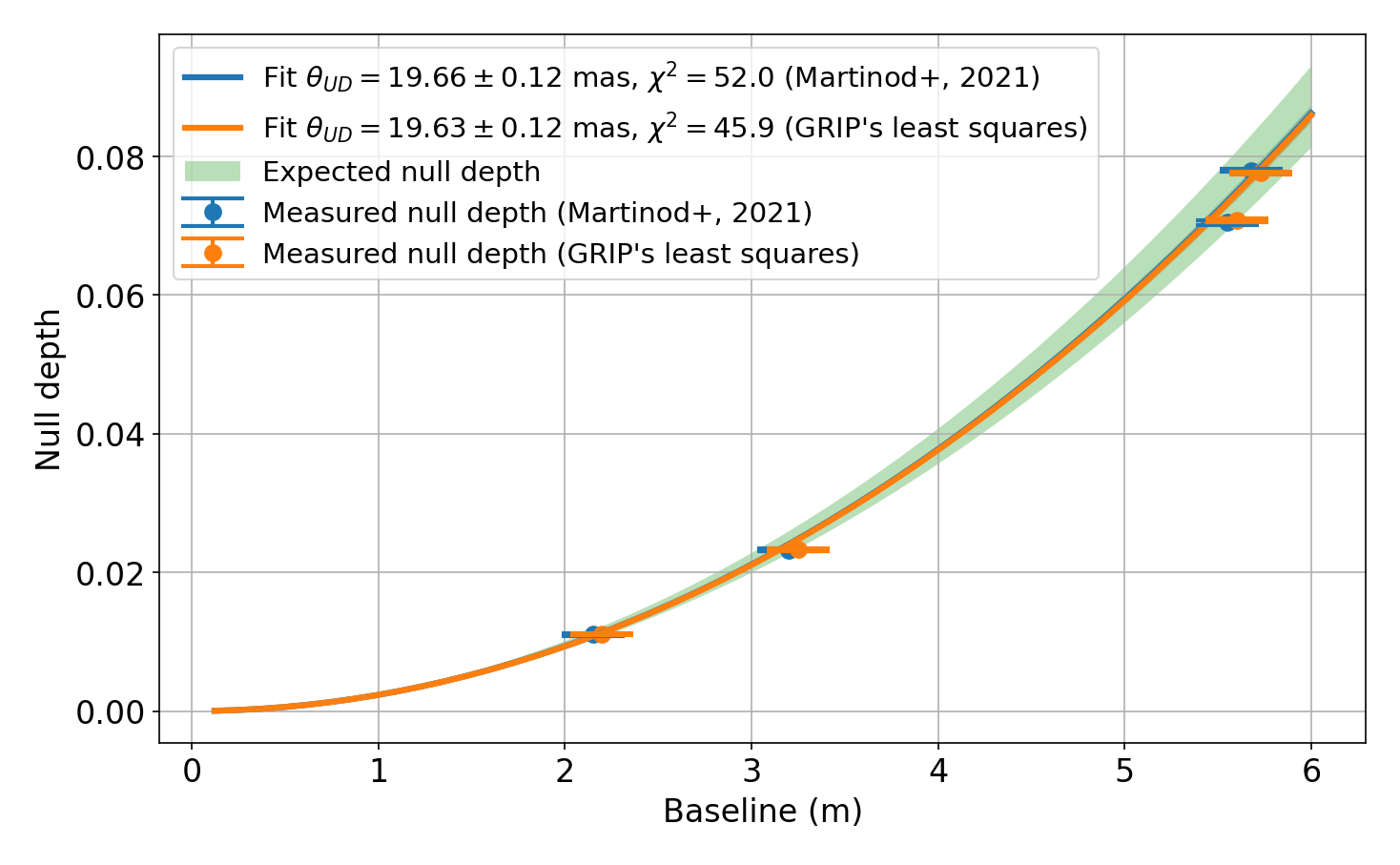} \\
        \includegraphics[width=0.75\textwidth]{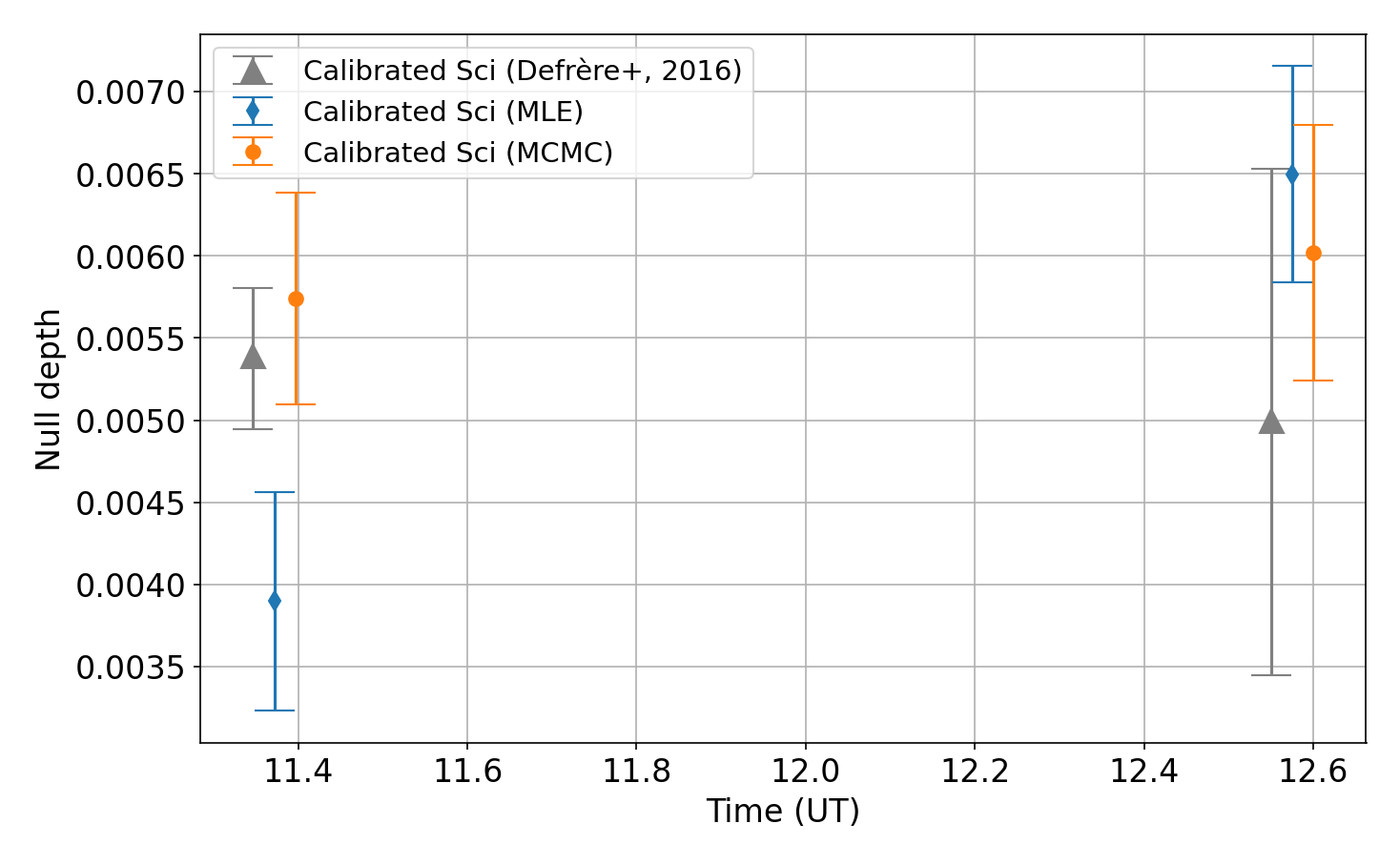} \\
    \end{tabular}
    \caption{Comparison of the results obtained with GRIP for the published data on GLINT (top) and LBTI (bottom) data. On each figure, the data points from \texttt{GRIP} are artificially horizontally shifted to clarify the plots.}
    \label{fig:final}
\end{figure}

We compare the calibrated null depths for both GLINT and LBTI cases with their published values.
Self-calibrated null depths from \texttt{GRIP} are completely consistent with the published values for GLINT and LBTI (Figure~\ref{fig:final}).
The null depths given by the package with the least squares method provide the same angular diameter of Arcturus on GLINT data.
For the LBTI case, the null depths measured on $\beta$~Leo, with the MLE and MCMC strategies, are calibrated by the transfer functions from the previous section.
The calibrated null depths on $\beta$~Leo obtained with \texttt{GRIP} with the MLE and MCMC methods are consistent with the published values.
Inddeed, the average null depth from the two $\beta$ Leo points \cite{defrere2016} is $0.52 \pm 0.013~$\% and $0.52 \pm 0.07~$\% from \texttt{GRIP} (the uncertainty is derived from the square root of the sum of the variances of each point).

\section{Conclusion}
\texttt{GRIP} is an open-source package providing methods to reduce data from any nuller according to the null self-calibration method.
This versatility is possible by providing a simulator of the of the instrumental perturbation on the observables along with the data, then using \texttt{GRIP} features to exploit these inputs.
This framework provides the basis for a standardised way of reducing nulling data, from nullers with different architectures or spectral coverage.
It has been proven to provide reliable results through several optimisation strategies and to be compatible with any kind of nuller.
\texttt{GRIP} is a scalable solution to implement new data reduction techniques, including neural network-based techniques or hierarchical Bayesian modelling. The results presented in this paper are consistent with previously published results and confirm interferometric nulling accuracy down to a few 10$^{-4}$. 
Future work will focus on pushing the nulling accuracy by implementing neural network-based techniques in \texttt{GRIP} and investigate its capabilities for stellar interferometers.

Besides, while \texttt{GRIP} adresses the question of reducing the data through the NSC approach to provide estimates of the null depth, it does not include their interpretation. 
This should be addressed in the future through the NIFITS standard, which will mirror the usage of OIFITS as a medium for astrophysical interpretation of the data. 
This format and the interface with \texttt{GRIP} is under investigation.

\section*{Acknowledgments}
M-A.M. has received funding from the European Union’s Horizon 2020 research and innovation programme under grant agreement No.\ 101004719.

D.D., G.G., and M-A.M. acknowledge support from the European Research Council (ERC) under the European Union's Horizon 2020 research and innovation program (grant agreement No.\ CoG - 866070).

R.L. has received funding from the Research Foundation -  Flanders (FWO) under the grant number 1234224N.

SE is supported by the National Aeronautics and Space Administration through the Astrophysics Decadal Survey Precursor Science program (Grant No. 80NSSC23K1473).

\section*{Data Availability}
The data produced in this study are available from the corresponding author on reasonable request.

\bibliography{article} 
\bibliographystyle{plainnat}

\end{document}